\documentstyle[aps,psfig,twocolumn]{revtex}
\begin{document}
\draft
\flushbottom
\twocolumn[
\hsize\textwidth\columnwidth\hsize\csname @twocolumnfalse\endcsname

\title{Properties of the ferrimagnetic 
double-perovskite A$_{2}$FeReO$_{6}$ (A=Ba
and Ca)}
\author{W. Prellier, V. Smolyaninova, Amlan Biswas, C. Galley and R.L. Greene}
\address{Center for Superconductivity Research, 
Department of Physics, University of\\
Maryland, College Park, MD 20742, USA}
\author{K. Ramesha and J. Gopalakrishnan}
\address{Solid State and Structural Chemistry 
Unit, India Institute of Science,\\
Bangalore 560012, India.}
\date{\today}
\maketitle
\tightenlines
\widetext
\advance\leftskip by 57pt
\advance\rightskip by 57pt

\begin{abstract}
Ceramics of A$_{2}$FeReO$_{6}$ double-perovskite have been prepared and
studied for A=Ba and Ca. Ba$_{2}$FeReO$_{6}$ has a cubic structure ($Fm3m$)
with $a\approx $8.0854(1) \AA\ whereas Ca$_{2}$FeReO$_{6}$ has a distorted
monoclinic symmetry with $a\approx $5.396(1) \AA , $b\approx $5.522(1) \AA ,\ 
$c\approx $7.688(2) \AA\ and $\beta $=90.4${{}^{\circ }}$ ($P21/n$). The
barium compound is metallic from 5 K to 385 K, i.e. no metal-insulator
transition has been seen up to 385 K, and the calcium compound is
semiconducting from 5 K to 385 K. Magnetization measurements show a
ferrimagnetic behavior for both materials, with T$_{c}$=315 K for Ba$_{2}$
FeReO$_{6}$ and above 385 K for Ca$_{2}$FeReO$_{6}$. A specific heat
measurement on the barium compound gave an electron density of states at the
Fermi level, $N(E_{F})$ equal to 6.1$\times 10^{24}$ eV$^{-1}$mole$^{-1}$.
At 5 K, we observed a negative magnetoresistance of 10 \% in a magnetic
field of 5 T, but only for Ba$_{2}$FeReO$_{6}$. Electrical, thermal and
magnetic properties are discussed and compared to the analogous compounds Sr$
_{2}$Fe(Mo,Re)O$_{6}$.
\end{abstract}

]
\narrowtext
\tightenlines

\section{Introduction}

Perovskite manganites exhibiting a variety of exotic electronic properties 
\cite{Millis} that include a spectacular decrease of electrical resistance
in a magnetic field \cite{VonHel}, the so-called colossal magnetoresistance
(CMR)\ \cite{Jin}, have attracted wide attention in recent years. A
significant feature of the electronic structure of the ferromagnetic CMR
manganites, revealed by recent experiments \cite{Park,Soulen} and theory 
\cite{pickett}, is that the charge carriers are almost completely
spin-polarized at the Fermi level E$_{F}$. These materials are half-metallic
ferromagnets, where the majority spin states near E$_{F}$ are delocalized
and the minority spin channel is effectively localized. Since half-metallic
ferromagnetism and magnetoresistance (MR), especially at low fields, seem to
be intimately related to each other \cite{Hwang} - the latter arising from
the former- there is an intense search for half-metallic magnets which could
be candidate materials for the realization of MR applications. While several
double-perovskite oxides of the kind A$_{2}$BB'O$_{6}$ (A being an alkaline
earth or rare earth ion and B, B' being $d$-transition metal ions) have been
theoretically predicted to be half-metallic antiferromagnets \cite{pickett2}
, one such material, Sr$_{2}$FeMoO$_{6}$ \cite{nakagawa}, has recently \cite
{Kobayashi,Kim} been shown to be a half-metallic ferrimagnet exhibiting a
significant tunneling-type magnetoresistance at room temperature. More
recently, Asano et al. \cite{asano} have shown that it is possible to have
either positive or negative MR in thin films of Sr$_{2}$FeMoO$_{6}$ grown by
pulsed laser deposition.

Ba$_{2}$FeReO$_{6}$ \cite{sleight} and Ca$_{2}$FeReO$_{6}$ are
double-perovskites whose structure and properties are quite similar to those
of Sr$_{2}$FeMoO$_{6}$. In both materials a valence degeneracy between the
B-site cation occurs, giving rise to the observed metallic and magnetic
properties ; in the SFMO case, the valence-degeneracy is between Fe$^{3+}$+Mo
$^{5+}\rightleftharpoons $ Fe$^{2+}$+Mo$^{6+}$ states, while in the Ba$_{2}$
FeReO$_{6}$ or Ca$_{2}$FeReO$_{6}$ \ case, the degenerate oxidation states
are Fe$^{3+}$+Re$^{5+}$ $\rightleftharpoons \ $Fe$^{2+}$+Re$^{6+}$. A major
difference between the Mo and Re oxides however, is that Mo$^{5+}$ is $
4d^{1} $, whereas Re$^{5+}$ is $5d^{2}$. This would mean that the conducting
charge carrier density of the Re compound would be twice as much as in the
compound Mo, while the localized spins centered on Fe remain the same in
both oxides. We believe that this difference could have an influence on the
magnetotransport behavior, especially in view of the recent report \cite
{majumdar} that the bulk low field MR in ferromagnetic metals is mainly
determined by the charge carrier density. In view of the foregoing, we
consider it important to investigate the magnetic and transport properties
of A$_{2}$FeReO$_{6}$ (A=Ba and Ca) and our results are reported in this
communication.

\section{Results and discussion}

Polycrystalline samples of A$_{2}$FeReO$_{6}$ (with\ A=Ba or Ca) were
synthesized by standard solid state methods. First, a precursor oxide of the
composition A$_{2}$ReO$_{5.5}$ (A=Ba or Ca) was prepared by reacting
stoichiometric amounts of ACO$_{3}$ and Re$_{2}$O$_{7}$ at 1000 ${^{\circ }}$
C in air for 2 h. Second, this resultant oxide was mixed with required
quantities of Fe$_{2}$O$_{3}$ and Fe powder to obtain the desired
composition A$_{2}$FeReO$_{6}$. Finally, pellets of this mixture were heated
in an evacuated sealed silica tube at 910 ${{}^{\circ }}$C during 4 days,
followed by another treatment at 960 ${{}^{\circ }}$C during the same time,
with intermediate grinding. X-Ray powder diffraction (XRD) patterns were
taken using a conventional diffractometer with Cu K$\alpha $ radiation ($
\lambda =$1.5406 \AA ). The XRD patterns of final compounds are shown in
Fig.1a and Fig.1b. Ba$_{2}$FeReO$_{6}$ is indexed on the basic of a cubic
cell ($Fm3m$) with $a\approx $8.054(1) \AA , similar to Ba$_{2}$YRuO$_{6}$ 
\cite{Battle}.\ Ca$_{2}$FeReO$_{6}$ is distorted to a monoclinic symmetry ($
P21/n$) with $a\approx $5.396(2) \AA , $b\approx $5.522(2) \AA , $c\approx $
7.688(1) \AA\ and $\beta =90.4{^{\circ }}${, similar to those of La}$_{{2}}${
CuIrO}$_{{6}}${\ and Nd}$_{{2}}${MgTiO}$_{{6}}${\ \cite{Anderson}.} These
results indicate the formation of the double-perovskite (Fig.2) for both
compounds, as reported earlier \cite{sleight} and consistent with recent
results \cite{Kobayashi,Kim}. One also notes that no impurity phase can be
detected in the XRD indicating a clean single phase in each case.
\begin{figure}
\centerline{
\psfig{figure=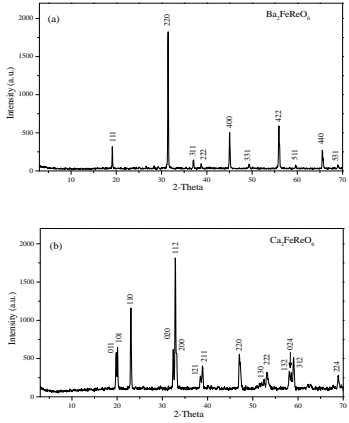,width=7.0cm,height=8.5cm,clip=}
}
\caption{X-Ray powder Diffraction of double-perovskite (a): Ba$_{2}$FeReO$_6$
,(b): Ca$_{2}$FeReO$_{6}$.}
\end{figure}
\begin{figure}
\centerline{
\psfig{figure=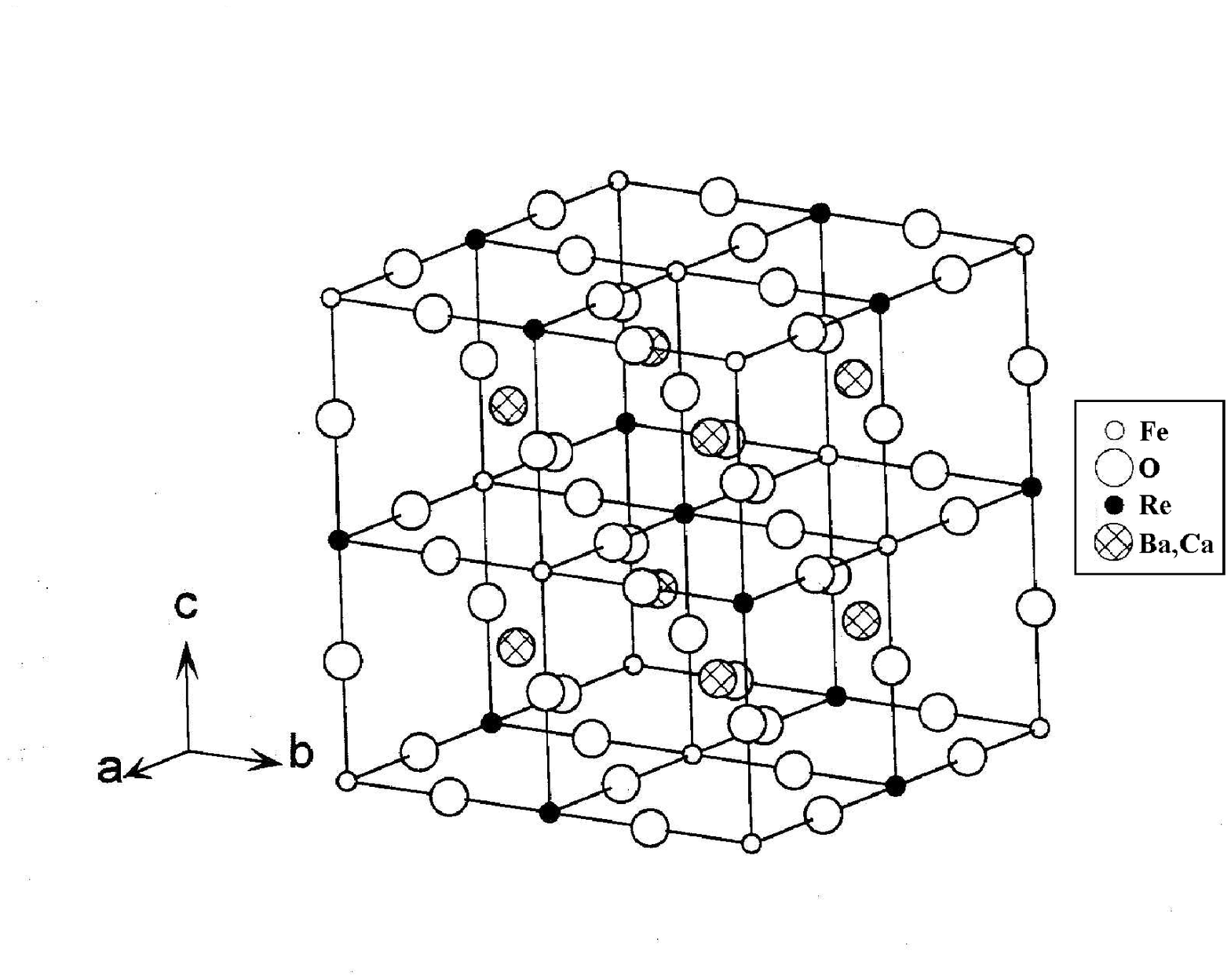,width=7.0cm,height=8.5cm,clip=}
}
\caption{Idealized structure of the ordered double-perovskite.}
\end{figure}

Resistivity ($\rho $) was measured between 5 K and 300 K, on bars with the
approximate dimensions of (1x3x8) mm$^{3}$, using the standard four-probe
method. The temperature (T) dependence of $\rho $ is shown in Fig.3 at
various magnetic fields of 0, 0.2 and 5 T for Ba$_{2}$FeReO$_{6}$. The
resistivity gradually decreases when the temperature decreases suggesting a
metallic behavior below 300 K. The zero-field resistivity was also measured
up to 385 K (inset of Fig.3) but no change in the metallic behavior was
observed in this region. In contrast, Ca$_{2}$FeReO$_{6}$ shows a
semiconducting behavior from 5 K to 300 K (Fig.4). While Ba$_{2}$FeReO$_{6}$
exhibits magnetoresistance at 5 K, the resistivity of Ca$_{2}$FeReO$_{6}$
remains unchanged, even under an applied magnetic field of 8 T. We find,
that the latter resistivity does not fit with an activation energy law ($
\rho (T)\propto e^{\frac{E_{a}}{kT}}$). As often found in the manganites\cite
{Urushibara}, the resistivity at room temperature is large (50 m$\Omega $.cm
for Ba$_{2}$FeReO$_{6}$ and 20 m$\Omega $.cm for Ca$_{2}$FeReO$_{6}$ ), well
above the Mott limit, but perhaps reflecting intergrain resistance in these
polycrystalline samples. A more detailed understanding of the transport of
these materials will require single crystals or oriented films.

To investigate the MR\ of the Ba$_{2}$FeReO$_{6}$, we present the MR at
different temperatures in Fig.5. As shown in Fig.3 , the Ba$_{2}$FeReO$_{6}$
compound exhibits negative MR, with MR defined as $MR(T,H)=\left[
R(H)-R(0)\right] /R(H)$. The MR is equal to 3 \% at 10 K in a 2000 Oe field.
This MR under low field at 10K is smaller than that found in Sr$_{2}$FeMoO$
_{6}$ (10 \%)\cite{Kobayashi}. However, the zero field $\rho (T)$ behavior
of Sr$_{2}$FeMoO$_{6}$ is rather different at low temperature than Ba$_{2}$
FeReO$_{6}$ -\ in Sr$_{2}$FeMoO$_{6}$\ it tends to increase slightly at 10
K. This behavior might be due to the preparation procedure of the sample
which dramatically affects the $\rho (T)$ \cite{Kobayashi}, most probably
this is a result of the difference in the structure of these compounds.
Indeed, in the series Ba$_{2}$FeReO$_{6}$, Sr$_{2}$FeReO$_{6}$ \cite{Abe}
and Ca$_{2}$FeReO$_{6}$, the crystal symmetry decreases from cubic to
tetragonal to orthorhombic (or distorted monoclinic). This is clearly a
manifestation of the decreasing Re-O-Fe bond angle from 180 degrees.
Accordingly, the conduction band width would be expected to decrease as we
go from Ba to Sr to Ca. Thus, it is not surprising that Ba$_{2}$FeReO$_{6}$
is metallic and Ca$_{2}$FeReO$_{6}$ is not. A similar conclusion occurs when
comparing Ba$_{2}$FeMoO$_{6}$ and Sr$_{2}$FeMoO$_{6}$ \cite
{Kobayashi,Maignan} ; Ba$_{2}$FeMoO$_{6}$ is metallic (and cubic) whereas Sr$
_{2}$FeMoO$_{6}$, whose structure is tetragonal, has a resistivity which
increases when T decreases \cite{Kobayashi}. The MR strongly increases at
low field with a slower increase at higher field. This effect occurs mainly
at low temperature, since at room temperature the MR\ is very small. The
features are characteristic of intergrain magnetoresistance \cite{Kim}. At
low temperature (Fig.5), a small hysteretic behavior also appears but thus
is not of relevance to the issues discussed in this paper.
\begin{figure}
\centerline{
\psfig{figure=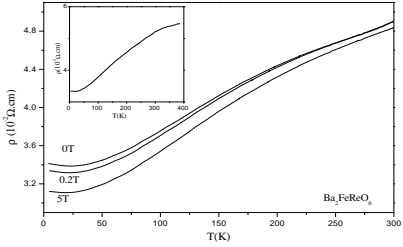,width=7.0cm,height=8.5cm,clip=}
}
\caption{Resistivity vs temperature under different magnetic fields for Ba$
_{2}$FeReO$_{6}$. The inset shows the zero field dependence of the
resistivity from 5-385 K.}
\end{figure}
\begin{figure}
\centerline{
\psfig{figure=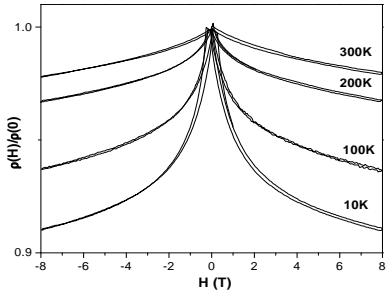,width=7.0cm,height=8.5cm,clip=}
}
\caption{Resistivity vs temperature under zero field for Ca$_{2}$ FeReO$_{6}$
.}
\end{figure}

Magnetization measurements (M)\ were made with a SQUID (MPMS Quantum Design)
magnetometer. These DC measurements were carried out with increasing
temperature after the sample was zero field cooled (ZFC). The temperature
dependence of M for a magnetic field of 10 Oe is shown in Fig.6 for Ba$_{2}$
FeReO$_{6}$. Like several other double-perovskite compounds \cite{Abe,Longo}
, A$_{2}$FeReO$_{6}$ (A=Ba and Ca) exhibit ferrimagnetic behavior due to an
antiferromagnetic superexchange interaction between spins of Fe$^{3+}$ ($
S=5/2$) and Re$^{5+}$ ($S=1$). The ferrimagnetic Curie temperature ($T_{c}$)
of Ba$_{2}$FeReO$_{6}$ was determined as 315 K from the temperature
dependence of the magnetization shown in Fig.6. For Ba$_{2}$FeReO$_{6}$ ,
the low temperature saturation magnetization value, taken under a magnetic
field of 2 T, is 24.9 emu/g, which is close to the expected value based on Fe
$^{3+}$ and Re$^{5+}$ moments (27.3 emu/g). The Ca$_{2}$FeReO$_{6}$ compound
exhibits a higher $T_{c}$, above 385 K (not determined). For the Ca$_{2}$
FeReO$_{6}$ compound, the low temperature saturation magnetization value,
measured under a field of 5 T, is calculated to be 40 emu/g which is a
little higher than the experimental value (30 emu/g) but in agreement with
a previous report \cite{Sleight2}. This is probably due to the fact that the
full saturation magnetization was not reached even with a 5T magnetic field.
The Curie temperature of Ba$_{2}$FeReO$_{6}$ is lower than the reported
value of T$_{c} \approx $410 K for Sr$_{2}$FeReO$_{6}$ \cite{Abe,Longo} but
close to the value of Ba$_{2}$FeMoO$_{6}$ ($T_{c}$=340 K) \cite{Maignan},
whereas the $T_{c}$ of Ca$_{2}$FeReO$_{6}$ seems to be higher ($>$ 385 K).
This means that these differences are likely caused by the larger ionic
radius of Ba$^{2+}$ (1.47 \AA\ versus 1.31 \AA\ for Sr$^{2+}$ and 1.18 \AA\
for Ca$^{2+}$ \cite{Shannon}) which leads to a larger Re-O-Fe (or Mo-O-Fe)
bond length and therefore a smaller exchange and lower Curie temperature.
\begin{figure}
\centerline{
\psfig{figure=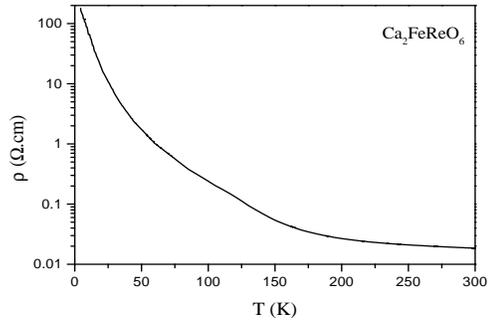,width=7.0cm,height=8.5cm,clip=}
}
\caption{Field dependence of the normalized MR at different temperatures for
Ba$_{2}$FeReO$_{6}$.}
\end{figure}
\begin{figure}
\centerline{
\psfig{figure=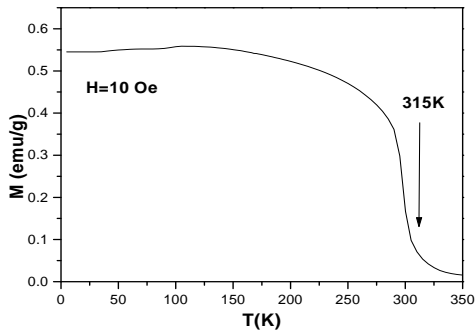,width=7.0cm,height=8.5cm,clip=}
}
\caption{Magnetization vs temperature under a magnetic field of 10 Oe for Ba$
_{2}$FeReO$_{6}$.}
\end{figure}

The specific heat was measured by relaxation calorimetry in the temperature
range 2-16 K. The specific heat data for Ba$_{2}$FeReO$_{6}$ and Ca$_{2}$
FeReO$_{6}$ are shown in Fig. 7. In our analysis of the low temperature
specific heat we include lattice ($C_{latt}$), metallic ($C_{el}$) and
hyperfine ($C_{hyp}$) contributions. When the temperature decreases below 3
K, the specific heat increases due to the hyperfine contribution $
C_{hyp}=A/T^{2}$. Since the nuclear spin of Fe$^{56}$ is zero, the hyperfine
contribution in our samples arises from the Re$^{186}$ nuclear spin $I=1$.
The experimental values of $A$ are found to be 135$\pm $4 mJ-Kmole and 180$
\pm $5 mJ-K/mole for Ba$_{2}$FeReO$_{6}$ and Ca$_{2}$FeReO$_{6}$
respectively. In addition, Ba$_{2}$FeReO$_{6}$ has the expected metallic
contribution, $\gamma T$, with $\gamma =23.1\pm 0.2$ mJ/mole-K$^{2}$. Using $
\gamma =\pi ^{2}k_{B}^{2}N(E_{F})/3$, we find the density of states at the
Fermi energy $N(E_{F})$ to be 6.1$\times 10^{24}$ eV$^{-1}$mole$^{-1}$. This
value is larger than the $N(E_{F})$ obtained from the band structure
calculation for Sr$_{2}$FeMoO$_{6}$ (1.2$\times 10^{24}$ eV$^{-1}$mole$^{-1}$
)\cite{Kobayashi}, probably because Re$^{5+}$ has two electrons on the $d$
orbital ($5d^{2}$), while Mo$^{5+}$ has only one ($4d^{1}$).

To achieve a good fit for our Ba$_{2}$FeReO$_{6}$ specific heat data two
lattice terms are required: $C_{latt}=\beta _{3}T^{3}+\beta _{5}T^{5}$,
where $\beta _{3}=0.438\pm 0.003$ mJ/mole-K$^{4}$ and $\beta _{5}=5.9\times
10^{-4}$ mJ/mole-K$^{6}$. Since $\Theta _{D}=(12\pi ^{4}pR/5\beta
_{3})^{1/3} $, where p=10 is the number of atoms per formula unit, we find
the Debye temperature $\Theta _{D}$ to be 354 K, which is similar to $\Theta
_{D}$ of other perovskites \cite{Hamilton}.

The best fit for Ca$_{2}$FeReO$_{6}$ does not include $\gamma T$ and $\beta
_{5}T^{5}$ terms, but requires an additional term $C^{\prime }\approx \alpha
T^{a}$, where $a$ is close to 2. The absence of the charge carrier term $
\gamma T$ is consistent with the insulating resistivity of Ca$_{2}$FeReO$
_{6} $. The $C^{\prime }\approx \alpha T^{a}$ term, or more precisely 
\begin{figure}
\centerline{
\psfig{figure=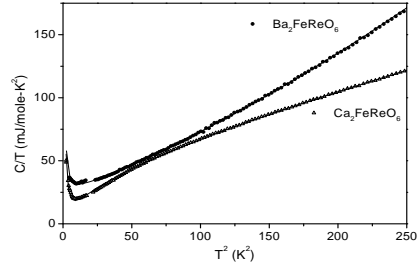,width=7.0cm,height=8.5cm,clip=}
}
\caption{Specific heat of Ba$_{2}$FeReO$_{6}$ and Ca$_{2}$FeReO$_{6}$
plotted as $C/T$ vs $T^{2}$. Lines are the best fit to the form $
C=A/T^{2}+\gamma T+\beta _{3}T^{3}+\beta _{5}T^{5}$ for Ba$_{2}$FeReO$_{6}$
and $C=A/T^{2}+\gamma T+\beta _{3}T^{3}+C^{\prime }$ for Ca$_{2}$FeReO$_{6}$
(see text for values parameters and definition of ${C}^{\prime}$).}
\end{figure}
$C^{\prime }=C^{\prime }(\Delta ,B,T)$, which is the specific heat of
excitations with a dispersion relation $\epsilon =\Delta +Bk^{2}$ of
non-magnetic origin, was also found in the charge-ordered perovskite
manganites \cite{Vera}. In the case of insulating Ca$_{2}$FeReO$_{6}$ a
charge of 3+ on the Fe site and a charge of 5+ on the Re site create a
situation similar to the Mn$^{3+}$-Mn$^{4+}$ charge ordering in manganites.
The fit to our Ca$_{2}$FeReO$_{6}$ specific heat data gives $\Delta $ = 7 K
and $B$ = 22 meV-\AA\ similar in magnitude to the parameters found in the
manganites. We believe that the reasons for the presence of the $C^{\prime
}(\Delta ,B,T)$ term in Ca$_{2}$FeReO$_{6}$ are similar to those in La$
_{0.5} $Ca$_{0.5}$MnO$_{3}$ \cite{Vera}. The lattice contribution $\beta
_{3}T^{3}$ ($\beta _{3}=0.25\pm 0.006$ mJ/mole-K$^{4}$) in Ca$_{2}$FeReO$_{6}
$ is smaller than in Ba$_{2}$FeReO$_{6} $, which can be explained by the
different crystal structure (cubic for Ba$_{2}$FeReO$_{6}$ and distorted
monoclinic for Ca$_{2}$FeReO$_{6}$) and smaller Ca mass.

A ferrimagnet has a magnetic contribution to the specific heat, $
C_{mag}=\delta T^{3/2}$, which is similar to that of a ferromagnetic one.
However, this contribution, can not be resolved from the specific heat data
alone, since our data can be fit well without the magnetic term. Since Ca$
_{2}$FeReO$_{6}$ and Ba$_{2}$FeReO$_{6}$ have a high Curie temperature, and
hence, strong exchange interaction (J), the magnetic term should be small
(since $\delta \propto \frac{1}{J^{3/2}}$).

\section{Conclusion}

In summary, we have investigated the transport, thermal and magnetic
properties of two polycrystalline double-perovskites Ba$_{2}$FeReO$_{6}$ and
Ca$_{2}$FeReO$_{6}$. Ba$_{2}$FeReO$_{6}$ displays a metallic behavior below
385 K whereas Ca$_{2}$FeREO$_{6}$ is insulating below this temperature. The
specific heat of Ba$_{2}$FeReO$_{6}$ gives a low temperature metallic
contribution with an electron density of states at the Fermi level close to
the band structure value. Insulating Ca$_{2}$FeReO$_{6}$ has no metallic
term in the specific heat but rather an extra contribution most likely
caused by charge ordering of Fe, Re. The Ba$_{2}$FeReO$_{6}$ compound
exhibits a negative MR at 10K, smaller than the analogous compound Sr$_{2}$
FeMoO$_{6}$. Magnetic measurements indicate a ferrimagnetic behavior,with a $
T_{c}=315K$ for Ba$_{2}$FeReO$_{6}$ and above 385 K for Ca$_{2}$FeReO$_{6}$
. These data have been explained and compared with the analogous compounds Sr
$_{2}$Fe(Re,Mo)O$_{6}$.

\acknowledgments{\ Partial support of NSF-MRSEC at University of Maryland is
acknowledged. The work at Bangalore was supported by the Department of
Science and Technology, Government of India. K.R. thanks the Council of
Scientific and Industrial Research, New Delhi, for the award of a fellowship.
}

\end{document}